\documentclass[12pt]{article}
\usepackage[latin1]{inputenc}
\usepackage[authoryear]{natbib}
\usepackage{graphicx}
\usepackage{amssymb, amsmath, amsthm}
\usepackage{times}
\usepackage{color}
\usepackage{bm}

\newcommand{\Es}{{\rm E}}
\newcommand{\diag}{{\rm diag}}
\newcommand{\e}{{\rm e}}
\newcommand{\dd}{{\rm d}}

\newcommand{\SLR}{S_{\textrm{LR}}}
\newcommand{\SW}{S_{\textrm{W}}}
\newcommand{\SR}{S_{\textrm{R}}}
\newcommand{\ST}{S_{\textrm{T}}}
\newcommand{\LR}{{\textrm{LR}}}
\newcommand{\W}{{\textrm{W}}}
\newcommand{\R}{{\textrm{R}}}
\newcommand{\T}{{\textrm{T}}}

\title{Improved likelihood inference in generalized linear models}
\author{Tiago M.~Vargas,\quad Silvia L.P.~Ferrari\\
{\small {\em Departamento de Estat\'istica, Universidade de S\~ao Paulo, S\~ao Paulo/SP, Brazil}}\\
Artur J.~Lemonte \\
{\small {\em Departamento de Estat\'istica, Universidade Federal de Pernambuco, Recife/PE, Brazil}}
}
\date{}

\begin{document}
\maketitle

\begin{abstract}

We address the issue of performing testing inference in
generalized linear models when the sample size is small.
This class of models provides a straightforward way of modeling normal
and non-normal data and has been widely used in several
practical situations. The likelihood ratio, Wald and score statistics,
and the recently proposed gradient statistic provide
the basis for testing inference on the parameters in these
models. We focus on the small-sample case, where the reference chi-squared
distribution gives a poor approximation to the true null distribution
of these test statistics. We derive a general Bartlett-type correction factor
in matrix notation for the gradient test which reduces the size distortion of the test,
and numerically compare the proposed test with the usual
likelihood ratio, Wald, score and gradient tests, and
with the Bartlett-corrected likelihood ratio and score tests.
Our simulation results suggest that the corrected test we propose
can be an interesting alternative to the other tests
since it leads to very accurate inference even for very small samples.
We also present an empirical application for illustrative purposes.
\footnote{Supplementary Material presents derivation of Bartlett-type 
corrections to the gradient tests, and the computer code used in Section \ref{application}.} \\

\end{abstract}

\section{Introduction}\label{introduction}

The likelihood ratio (LR), Wald and Rao score tests
are the large-sample tests usually employed for
testing hypotheses in parametric models.
Another criterion for testing hypotheses in parametric models,
referred to as the  gradient test, was proposed by \cite{Terrell2002}, and has received considerable
attention in the last few years. An advantage of the gradient statistic over the Wald and the
score statistics is that it does not involve knowledge
of the information matrix, neither expected nor observed. Additionally,
the gradient statistic is quite simple to be computed. Here, it is worthwhile to quote
\cite{Rao2005}: ``The suggestion by Terrell is attractive as it is simple to compute.
It would be of interest to investigate the performance of the [gradient] statistic.''
Also, Terrell's statistic shares the same first
order asymptotic properties with the LR, Wald and score statistics.
That is, to the first order of approximation, the LR,
Wald, score and gradient statistics have the same asymptotic
distributional properties either under the null hypothesis or under a sequence of
Pitman alternatives, i.e.~a sequence of local
alternative hypotheses that shrink to the null hypothesis at a convergence rate $n^{-1/2}$,
$n$ being the sample size; see \cite{LemonteFerrari2012a}.

The LR, Wald, score and gradient statistics for testing composite or simple null
hypothesis $\mathcal{H}_{0}$ against an alternative hypothesis $\mathcal{H}_{a}$,
in regular problems, have a $\chi^2_k$ null distribution asymptotically,
where $k$ is the difference between
the dimensions of the parameter spaces under the two hypotheses being tested.
However, in small samples, the use of these statistics coupled with their asymptotic properties 
become less justifiable. One way of improving the $\chi^2$ approximation
for the exact distribution of the LR statistic is by multiplying it by a correction factor known as the
Bartlett correction \citep[see][]{Bartlett37}. This idea was later put into a
general framework by \citet{Lawley1956}. The $\chi^2$ approximation
for the exact distribution of the score statistic can be
improved by multiplying it by a correction factor known as the
Bartlett-type correction. It was demonstrated in a general framework by
\cite{CordeiroFerrari1991}. Recently, \cite{VargasFerrariLemonte2013} demonstrated
how to improve the $\chi^2$ approximation for the exact distribution
of the gradient statistic in 
wide generality by multiplying it by a Bartlett-type correction factor. There is no
Bartlett-type correction factor to improve the $\chi^2$ approximation of the
exact distribution of the Wald statistic in a general
setting. The Bartlett and Bartlett-type corrections became widely used
for improving the large-sample $\chi^2$ approximation
to the null distribution of the LR and score statistics in
several special parametric models. In recent years there has
been a renewed interest in Bartlett and Bartlett-type factors and several
papers have been published giving expressions for computing
these corrections for special models. Some references are
\cite{Zucker-et-al-2000}, \cite{ManorZucker2004}, \cite{LagosMorettin2004}, \cite{Tu-et-al-2005},
\cite{vanGiersbergen2009}, \cite{Bai-2009}, \cite{Lagos-et-al-2010},
\cite{Lemonte-et-al-2010}, \cite{LemonteFerrari2011}, \cite{Noma-2011}, \cite{Fujita-et-al-2010}, \cite{BayerCribari2012},
\cite{Lemonte-et-al-2012}, among others.
The reader is referred to \cite{CordeiroCribari1996}
for a detailed survey on Bartlett and Bartlett-type corrections.

The generalized linear models (GLMs), first defined by \cite{NelderWedderburn1972},
are a large class of statistical models for relating responses to linear combinations of
predictor variables, including many commonly encountered
types of dependent variables and error structures as special cases. It
generalizes the classical normal linear model,
by relaxing some of its restrictive assumptions, and provides methods
for the analysis of non-normal data. Additionally, the GLMs have applications
in disciplines as widely varied as agriculture, demography, ecology,
economics, education, engineering, environmental studies,
geography, geology, history, medicine, political science, psychology,
and sociology. We refer the read to \cite{Lindsey1997} for applications of GLMs
in these areas. In summary,  the GLM approach is
attractive because it (1) provides a general theoretical
framework for many commonly encountered statistical models; (2) simplifies
the implementation of these different models in statistical
software, since essentially the same algorithm can be used
for estimation, inference and assessing model adequacy for all GLMs.
Introductions to the area are given by \cite{Firth1991} and \cite{DobsonBarnett2008}, whereas
\cite{McCullaghNelder1989} and \cite{HardinHilbe2007} give more comprehensive treatments.

The asymptotic $\chi^2$ distribution of the LR, Wald, score and gradient statistics is
used to test hypotheses on the model parameters in GLMs, since their exact
distributions are difficult to obtain in finite samples. However, for
small sample sizes, the $\chi^2$ distribution may not be a
trustworthy approximation to the exact null distributions of
the LR, Wald, score and gradient statistics.
Higher order asymptotic methods, such as the Bartlett and Bartlett-type
corrections, can be used to improve the LR, Wald, score and gradient tests. 
Several papers have focused on deriving matrix formulas for the Bartlett
and Bartlett-type correction factors in GLMs.
For example, some efforts can be found in the works by \cite{Cordeiro83, Cordeiro87},
who derived an improved LR statistic. An improved
score statistic was derived by \cite{CordeiroFerrariPaula1993}
and \cite{CribariFerrari1995}. These results will be revised in
this paper. Although the algebraic forms of the Bartlett and Bartlett-type correction factors
are somewhat complicated, they can be easily incorporated into a computer program. This might
be a worthwhile practice, since the Bartlett and Bartlett-type
corrections act always in the right direction and, in general, give a substantial improvement.

This paper is concerned with small sample likelihood inference in
GLMs. First, we derive a general Bartlett-type correction factor in matrix
notation to improve the inference based on the gradient statistic
in the class of GLMs when the number
of observations available to the practitioner is small.
Further, in order to evaluate and compare the finite-sample performance of the improved
gradient test in GLMs with the usual LR, Wald, score and
gradient tests, and with the improved LR and score tests,
we also perform Monte Carlo simulation experiments by considering the gamma regression model and
the inverse normal regression model. The simulation study on the
size properties of these tests evidences that the improved gradient
test proposed in this paper can be an appealing alternative to the
classic asymptotic tests in this class of models when the number of
observations is small. We shall emphasize that we have not found any comprehensive 
simulation study in the statistical literature comparing the classical uncorrected and corrected 
large-sample tests in GLMs. This paper fills this gap, and includes the gradient test and 
its Bartlett-type corrected version derived here in the simulation study.

The article is organized in the following form.
In Section \ref{GMLs-sec:2}, we define
the class of GLMs and discuss estimation and hypothesis testing inference on the regression
parameters. Improved likelihood-based
inference is presented in Section \ref{improveTests}. We
present the Bartlett-corrected LR and score statistics, and
derive a Bartlett-type correction factor for the gradient
statistic. Tests on the precision parameter are provided in Section \ref{phi-tests}.
Monte Carlo simulation results are presented and discussed in Section \ref{simulations}.
An application to real data is considered in Section \ref{application}. The
paper closes up with a brief discussion in Section \ref{conclusion}.

\section{The model, estimation and testing}\label{GMLs-sec:2}

Suppose the univariate random variables $Y_1,\ldots,Y_n$ are independent and each
$Y_l$ has a probability density function in the following family of
distributions:
\begin{equation}\label{dens1}
\pi(y; \theta_{l},\phi)=\exp\bigl\{\phi[y\theta_{l} - b(\theta_{l}) + c(y)] + a(y,\phi)\bigr\},\qquad l=1,\ldots,n,
\end{equation}
where $a(\cdot,\cdot)$, $b(\cdot)$ and $c(\cdot)$ are known appropriate functions.
The mean and the variance of $Y_l$ are
$\Es(Y_{l}) = \mu_{l} = \dd b(\theta_{l})/\dd\theta_{l}$ and
var$(Y_{l}) = \phi_{l}^{-1}V_{l}$, where
$V_l = \dd\mu_{l}/\dd\theta_{l}$ is called the variance function
and $\theta_{l}=q(\mu_{l})=\int V_{l}^{-1}\dd\mu_{l}$ is a known
one-to-one function of $\mu_{l}$. The choice of the variance function
$V_{l}$ as a function of $\mu_{l}$ determines $q(\mu_{l})$. We have
$V_{l}=1\,[q(\mu_{l})=\mu_{l}]$, $V_{l}=\mu_{l}^2\,[q(\mu_{l})=-1/\mu_{l}]$
and $V_{l}=\mu_{l}^3\,[q(\mu_{l})=-1/(2\mu_{l}^2)]$ for the
normal, gamma and inverse normal distributions, respectively.
The parameters $\theta_{l}$ and $\phi>0$ in (\ref{dens1}) are called
the canonical and precision parameters, respectively, and the inverse
of $\phi$, $\phi^{-1}$, is the dispersion parameter of the distribution.

In order to introduce a regression structure in the class of models in \eqref{dens1},
we assume that
\begin{equation}\label{sist_part}
d(\mu_{l}) = \eta_{l} = \bm{x}_{l}^\top\bm{\beta},
\qquad l = 1,\ldots,n,
\end{equation}
where $d(\cdot)$ is a known one-to-one differentiable link
function, $\bm{x}_{l}^{\top} = (x_{l1},\ldots,x_{lp})$
is a vector of known variables associated with the $l$th observable
response, and $\bm{\beta}=(\beta_{1},\ldots,\beta_{p})^\top$ is a set
of unknown parameters to be estimated from the data ($p <n$).
The regression structure links the covariates $\bm{x}_l$ to the parameter
of interest $\mu_{l}$. Here, we assume only identifiability
in the sense that distinct $\beta$'s imply distinct $\eta$'s.
Further, the precision parameter may be known or unknown, and it is the same 
for all observations.

Let $\ell(\bm{\beta},\phi)$ denote the total log-likelihood function for a given GLM.
We have
\[
\ell(\bm{\beta},\phi)=\phi\sum_{l=1}^{n}[y_l \theta_{l} - b(\theta_l)+c(y_l)] + \sum_{l=1}^{n}a(y_l,\phi),
\]
where $\theta_{l}$ is related to $\bm{\beta}$ by (\ref{sist_part}).
The total score function and the total Fisher information matrix for $\bm{\beta}$
are given, respectively, by
$\bm{U}_{\bm{\beta}}=\phi\bm{X}^{\top}\bm{W}^{1/2}\bm{V}^{-1/2}(\bm{y}-\bm{\mu})$ and
$\bm{K}_{\bm{\beta}} = \phi\bm{X}^{\top}\bm{W}\bm{X}$,
where $\bm{W}=\diag\{w_{1},\ldots,w_{n}\}$ with $w_{l} = V_{l}^{-1}(\dd\mu_{l}/\dd\eta_{l})^2$,
$\bm{V}=\diag\{V_{1},\ldots,V_n\}$, $\bm{y}=(y_1,\ldots,y_n)^\top$ and
$\bm{\mu}=(\mu_{1},\ldots,\mu_{n})^\top$. The model
matrix $\bm{X}=(\bm{x}_1,\ldots,\bm{x}_n)^\top$ is assumed to be of full rank, i.e.~rank$(\bm{X})=p$.
The maximum likelihood estimate
(MLE) $\widehat{\bm{\beta}}$ of $\bm{\beta}$ can be obtained iteratively using the
standard reweighted least squares method
\[
\bm{X}^{(m)\top}\bm{W}^{(m)}\bm{X}^{(m)}\bm{\beta}^{(m+1)} = \bm{X}^{(m)\top}\bm{W}^{(m)}\bm{y}^{*(m)},
\qquad m = 0,1,\ldots,
\]
where $\bm{y}^{*(m)} = \bm{X}^{(m)}\bm{\beta}^{(m)} + \bm{N}^{(m)}(\bm{y}-\bm{\mu}^{(m)})$
is a modified dependent variable and the matrix $\bm{N}$ assumes the form
$\bm{N}=\diag\{(\dd\mu_{1}/\dd\eta_{1})^{-1}, \ldots, (\dd\mu_{n}/\dd\eta_{n})^{-1}\}$.
The above equation shows that any software with a weighted regression routine can be used
to evaluate $\widehat{\bm{\beta}}$. Additionally, note that 
$\widehat{\bm{\beta}}$ does not depend on the parameter $\phi$.

Estimation of the dispersion parameter $\phi$ by the maximum likelihood method
is a more difficult problem than the estimation of $\bm{\beta}$ and the complexity
depends on the functional form of $a(y,\phi)$. The MLE $\widehat{\phi}$
of $\phi$ is a function of the deviance ($D_p$) of the model,
which is defined as $D_p = 2\sum_{l=1}^{n}[v(y_{l})-v(\widehat{\mu}_l)
+ (\widehat{\mu}_l - y_{l})q(\widehat{\mu}_{l})]$, where
$v(z) = zq(z) - b(q(z))$ and $\widehat{\mu}_l$
denotes the MLE of $\mu_{l}$ ($l=1,\ldots,n$). That is, given the estimate
$\widehat{\bm{\beta}}$, the MLE of $\phi$ can be found as the solution of the equation
\[
\sum_{l=1}^{n}\frac{\partial a(y_{l},\phi)}{\partial\phi}\biggr|_{\phi=\widehat{\phi}}
=\frac{D_{p}}{2} - \sum_{l=1}^{n}v(y_{l}).
\]
When (\ref{dens1}) is a two-parameter full exponential family distribution with canonical
parameters $\phi$ and $\phi\theta$, the term $a(y,\phi)$ in \eqref{dens1}
can be expressed as $a(y,\phi) = \phi a_0(y) + a_1(\phi) + a_2(y)$,
and the estimate of $\phi$ is obtained from
\[
a'_{1}(\widehat{\phi}) = \frac{1}{n}\biggl[\frac{D_{p}}{2} - \sum_{l=1}^{n}t(y_{l})\biggr],
\]
where $a'_{1}(\phi) = \dd a_1(\phi)/\dd\phi$ and $t(y_{l}) = v(y_{l}) + a_{0}(y_{l})$,
for $l=1,\ldots,n$. Table \ref{tab1:1} lists the functions $a_1(\phi)$, $v(y)$ and $t(y)$
for normal, inverse normal and gamma models. For normal and
inverse normal models we have that $\widehat{\phi} = n/D_p$, whereas
for the gamma model the MLE $\widehat{\phi}$ is obtained from $\log(\widehat{\phi}) -
\psi(\widehat{\phi}) = D_{p}/(2n)$, where $\psi(\cdot)$ is the digamma function, thus requiring the
use of a nonlinear numerical algorithm. For further details see \cite{cordMcCulagh1991}.
\begin{table}[!htp]
\begin{center}
\caption{Some special models.$^a$}\label{tab1:1}
\begin{tabular}{lccc}\hline
Model            &  $a_1(\phi)$                             & $v(y)$      &  $t(y)$ \\\hline
Normal           &  $\log(\phi)/2$                          & $y^2/2$     &  0    \\
Inverse normal   &  $\log(\phi)/2$                          & $1/(2y)$    &  0    \\
Gamma            &  $\phi\log(\phi) - \log\{\Gamma(\phi)\}$ & $\log(y)-1$ &  $-1$ \\\hline
\multicolumn{4}{l}{{\small $^a$$\Gamma(\cdot)$ is the gamma function.}}
\end{tabular}
\end{center}
\end{table}

In the following, we shall consider the tests which are based on the
LR ($S_{\LR}$), Wald ($S_\W$), Rao score ($S_\R$) and gradient ($S_\T$) statistics
in the class of GLMs for testing a composite null hypothesis.
The hypothesis of interest is $\mathcal{H}_{0}: \bm{\beta}_{1} = \bm{\beta}_{10}$,
which will be tested against the alternative hypothesis
$\mathcal{H}_{a}:\bm{\beta}_{1}\neq\bm{\beta}_{10}$,
where $\bm{\beta}$ is partitioned as $\bm{\beta} = (\bm{\beta}_{1}^{\top},
\bm{\beta}_{2}^{\top})^{\top}$, with
$\bm{\beta}_{1} = (\beta_{1}, \dots,\beta_{q})^{\top}$ and
$\bm{\beta}_{2} = (\beta_{q+1},\dots,\beta_{p})^{\top}$. Here,
$\bm{\beta}_{10}$ is a fixed column vector of dimension $q$, and
$\bm{\beta}_2$ and $\phi$ act as nuisance parameters. The
partition of the parameter vector $\bm{\beta}$ induces the corresponding partitions:
$\bm{U}_{\bm{\beta}} = (\bm{U}_{\bm{\beta}_1}^\top, \bm{U}_{\bm{\beta}_2}^\top)^\top$,
with $\bm{U}_{\bm{\beta}_1}=\phi\bm{X}_1^{\top}\bm{W}^{1/2}\bm{V}^{-1/2}(\bm{y}-\bm{\mu})$ and
$\bm{U}_{\bm{\beta}_2}=\phi\bm{X}_2^{\top}\bm{W}^{1/2}\bm{V}^{-1/2}(\bm{y}-\bm{\mu})$,
\[
\bm{K}_{\bm{\beta}} =
\begin{bmatrix}
\bm{K}_{\bm{\beta}11} & \bm{K}_{\bm{\beta}12} \\
\bm{K}_{\bm{\beta}21} & \bm{K}_{\bm{\beta}22}
\end{bmatrix} = \phi
\begin{bmatrix}
\bm{X}_1^{\top}\bm{W}\bm{X}_1 & \bm{X}_1^{\top}\bm{W}\bm{X}_2 \\
\bm{X}_2^{\top}\bm{W}\bm{X}_1 & \bm{X}_2^{\top}\bm{W}\bm{X}_2
\end{bmatrix},
\]
with the matrix $\bm{X}$ partitioned as $\bm{X} = \bigl[\bm{X}_{1}\ \ \bm{X}_{2}\bigr]$,
$\bm{X}_{1}$ being $n\times q$ and $\bm{X}_{2}$ being $n\times (p-q)$.
The LR, score, Wald and gradient statistics for testing $\mathcal{H}_{0}$
can be expressed, respectively, as
\[
\SLR = 2\bigl[\ell(\widehat{\bm{\beta}}_{1}, \widehat{\bm{\beta}}_{2},\widehat{\phi})
- \ell(\bm{\beta}_{10}, \widetilde{\bm{\beta}}_{2},\widetilde{\phi})\bigr],
\]
\[
\SR = \widetilde{\bm{s}}^{\top}\widetilde{\bm{W}}^{1/2}\bm{X}_1(\widetilde{\bm{R}}^{\top}\widetilde{\bm{W}}\widetilde{\bm{R}})^{-1}
\bm{X}_{1}^{\top}\widetilde{\bm{W}}^{1/2}\widetilde{\bm{s}},
\]
\[
\SW = \widehat{\phi}(\widehat{\bm{\beta}}_1 - \bm{\beta}_{10})^{\top}
(\widehat{\bm{R}}^{\top}\widehat{\bm{W}}\widehat{\bm{R}})(\widehat{\bm{\beta}}_1 - \bm{\beta}_{10}),
\]
\[
\ST = \widetilde{\phi}^{1/2}\widetilde{\bm{s}}^{\top}\widetilde{\bm{W}}^{1/2}\bm{X}_1(\widehat{\bm{\beta}}_1 - \bm{\beta}_{10}),
\]
where $(\widehat{\bm{\beta}}_{1}, \widehat{\bm{\beta}}_{2}, \widehat{\phi})$ and
$(\bm{\beta}_{10}, \widetilde{\bm{\beta}}_{2},\widetilde{\phi})$ are the unrestricted
and restricted (under $\mathcal{H}_{0}$) MLEs of $(\bm{\beta}_{1}, \bm{\beta}_{2}, \phi)$, respectively,
$\bm{s} = \phi^{1/2}\bm{V}^{-1/2}(\bm{y}-\bm{\mu})$ is the Pearson residual vector and
$\bm{R} = \bm{X}_{1} - \bm{X}_{2}\bm{A}$, where
$\bm{A} = (\bm{X}_{2}^{\top}\bm{W}\bm{X}_{2})^{-1}\bm{X}_{2}^{\top}\bm{W}\bm{X}_{1}$ represents a $(p-q)\times q$
matrix whose columns are the vectors of regression coefficients
obtained in the weighted normal linear regression of the columns of
$\bm{X}_1$ on the model matrix $\bm{X}_2$ with
$\bm{W}$ as a weight matrix. Here, tildes and hats indicate
quantities available at the restricted and unrestricted MLEs,
respectively. Under the null hypothesis $\mathcal{H}_{0}$,
these statistics have a $\chi_{q}^2$ distribution up to
an error of order $n^{-1}$.

\section{Improved inference in GLMs}\label{improveTests}

The chi-squared distribution may be a poor approximation to the null
distribution of the statistics discussed in Section \ref{GMLs-sec:2} when the sample size
is not sufficiently large. It is thus important
to obtain refinements for inference based on these tests from second-order
asymptotic theory. For GLMs, Bartlett correction factors for the LR statistic
were obtained by \cite{Cordeiro83, Cordeiro87}. Bartlett-type correction
factors in GLMs for the Rao score statistic were obtained by \cite{CordeiroFerrariPaula1993}
and \cite{CribariFerrari1995}. In addition to the corrected LR and score statistics
to test hypotheses on the model parameters in the class of GLMs,
we shall derive Bartlett-type correction factors for the gradient statistic in GLMs
on the basis of the general results in \cite{VargasFerrariLemonte2013}.
These results are new and represent additional contributions to improve
likelihood-based inference in GLMs.

To define the corrected LR and score statistics as well as to derive
Bartlett-type correction factors for the gradient statistic in GLMs,
some additional notation is in order. We define the matrices
$\bm{Z}=\bm{X}(\bm{X}^{\top}\bm{W} \bm{X})^{-1}\bm{X}^{\top}=((z_{lc}))$,
$\bm{Z}_2=\bm{X}_2(\bm{X}_2^{\top}\bm{W}\bm{X}_2)^{-1}\bm{X}_2^{\top}=((z_{2lc}))$,
$\bm{Z}_d=\diag\{z_{11},\ldots,z_{nn}\}$, $\bm{Z}_{2d}=\diag\{z_{211},\ldots,z_{2nn}\}$,
$\bm{F}=\diag\{f_1,\ldots,f_n\}$, $\bm{G}=\diag\{g_1,\ldots,g_n\}$,
$\bm{T}=\diag\{t_1,\ldots,t_n\}$, $\bm{D}=\diag\{d_1,\ldots,d_n\}$, $\bm{E}=\diag\{e_1,\ldots,e_n\}$,
$\bm{M}=\diag\{m_1,\ldots,m_n\}$,
$\bm{B} = \diag\{b_1,\ldots,b_n\}$, $\bm{H} = \diag\{h_1,\ldots,h_n\}$,
where
\[
f_l={\frac{1}{V_l}\frac{{\rm d}\mu_{l}}{{\rm d}\eta_{l}}\frac{{\rm d}^2 \mu_{l}}{{\rm d}\eta_{l}^2}},\qquad
g_l=f_l-\frac{1}{V_l^2}\frac{{\rm d}V_l}{{\rm d}\mu_{l}}\left(\frac{{\rm d}\mu_{l}}{{\rm d}\eta_{l}}\right)^3,
\]
\[
\lambda_{1l}=\frac{1}{V_l^2}\frac{{\rm d}V_l}{{\rm d}\mu_{l}}
\left(\frac{{\rm d}\mu_{l}}{{\rm d}\eta_{l}}\right)^2\frac{{\rm d}^2\mu_{l}}{{\rm d}\eta_{l}^2},
\qquad
\lambda_{2l}=\frac{1}{V_{l}}\left(\frac{{\rm d}^2\mu_{l}}{{\rm d}\eta_{l}^2}\right)^2,
\]
\[
\lambda_{3l}=\frac{1}{V_{l}}\frac{{\rm d}\mu_{l}}{{\rm d}\eta_{l}}\frac{{\rm d}^3\mu_{l}}{{\rm d}\eta_{l}^3},
\qquad
\lambda_{4l}= \frac{1}{V_{l}^3}\left(\frac{{\rm d}V_{l}}{{\rm d}\mu_{l}}\right)^2\left(\frac{{\rm d}\mu_{l}}{{\rm d}\eta_{l}}\right)^4,
\qquad
\lambda_{5l}= \frac{1}{V_{l}^2}\frac{{\rm d}^2 V_{l}}{{\rm d}\mu_{l}^2}\left(\frac{{\rm d}\mu_{l}}{{\rm d}\eta_{l}}\right)^4,
\]
\[
t_l=-9\lambda_{1l}+3\lambda_{2l}+3\lambda_{3l}+4\lambda_{4l}-2\lambda_{5l}, \qquad
d_l=-5\lambda_{1l}+2\lambda_{2l}+2\lambda_{3l}+2\lambda_{4l}-\lambda_{5l},
\]
\[
e_l=-12\lambda_{1l}+3\lambda_{2l}+4\lambda_{3l}+6\lambda_{4l}-3\lambda_{5l},
\]
\[
b_l=\lambda_{3l}+\lambda_{4l},
\qquad
h_l=\lambda_{1l}+\lambda_{5l},
\qquad
m_l=-4\lambda_{1l}+\lambda_{2l}+2\lambda_{4l}-\lambda_{5l}.
\]
We also define $\bm{Z}^{(2)}=\bm{Z}\otimes \bm{Z}$, $\bm{Z}_2^{(2)}=\bm{Z}_2\otimes \bm{Z}_2$,
$\bm{Z}^{(3)}=\bm{Z}^{(2)}\otimes \bm{Z}$, etc.,
where ``$\otimes$'' denotes the Hadamard (elementwise) product of matrices.
Let $\bm{1}_n=(1,\ldots,1)^\top$ be the $n$-vector of ones.
The matrices $\phi^{-1}\bm{Z}$ and $\phi^{-1}\bm{Z}_2$ have simple
interpretations as asymptotic covariance structures of
$\bm{X}\widehat{\bm{\beta}}$ and $\bm{X}_2\widetilde{\bm{\beta}}_2$, respectively.

From the general result of \cite{Lawley1956}, \cite{Cordeiro83, Cordeiro87}
defined the Bartlett-corrected LR statistic
for testing $\mathcal{H}_{0}: \bm{\beta}_{1} = \bm{\beta}_{10}$ in GLMs as
\begin{equation} \label{LRcor}
{S}^{*}_{\LR}=\frac{{S}_{\LR}}{1+a_{\LR}},
\end{equation}
where $a_{\LR}=A_{\LR}/(12q)$, $A_{\LR}=A_{\LR}+A_{\LR,\beta\phi}$,
\begin{align*}\label{alr}
A_{\LR}&=-4\phi^{-1}\bm{1}_n^{\top}\bm{G}(\bm{Z}^{(3)}-\bm{Z}^{(3)}_2)(\bm{F}+\bm{G})\bm{1}_n
+ 3\phi^{-1}\bm{1}_n^{\top}\bm{M}(\bm{Z}_d^{(2)}-\bm{Z}_{2d}^{(2)})\bm{1}_n\\
&\quad+ \phi^{-1}\bm{1}_n^{\top}\bm{F}[2(\bm{Z}^{(3)}-\bm{Z}^{(3)}_2)
+3(\bm{Z}_d\bm{Z}\bm{Z}_d - \bm{Z}_{2d}\bm{Z}_2\bm{Z}_{2d})]\bm{F}\bm{1}_n,
\end{align*}
\[
A_{\LR,\beta\phi}=\frac{3q}{nd_{(2)}^{2}}[d_{(2)}(2+q-2p)+2d_{(3)}],
\]
where $d_{(2)}=d_{(2)}(\phi)=\phi^2a_1''(\phi)$ and $d_{(3)}=d_{(3)}(\phi)=\phi^3a_1'''(\phi)$,
with $a_1''(\phi)={\rm d} a_1'(\phi)/{\rm d}\phi$ and  $a_1'''(\phi)={\rm d} a_1''(\phi)/{\rm d}\phi$;
when $\phi$ is known $A_{\LR,\beta\phi}$ is zero.  
The correction factor $1+a_{\LR}$ is commonly
referred to as the `Bartlett correction factor'.
It is possible to achieve a better $\chi_{q}^{2}$ approximation by using
the modified test statistic ${S}^{*}_{\LR}$ instead of ${S}_{\LR}$.
The adjusted statistic ${S}^{*}_{\LR}$ is $\chi_{q}^{2}$
distributed up to an error of order $n^{-2}$ under ${\mathcal H}_{0}$.

The Bartlett-corrected score statistic for testing $\mathcal{H}_{0}: \bm{\beta}_{1} = \bm{\beta}_{10}$ in GLMs
was derived in \cite{CordeiroFerrariPaula1993} and \cite{CribariFerrari1995}. It was obtained
by using the general result of \cite{CordeiroFerrari1991}. The corrected score
statistic is
\begin{align} \label{escorecorr}
{S}_{\R}^{*}={S}_{\R}\bigl[1-\bigl({c}_{\R}+{b}_{\R}{S}_\R+{a}_{\R}{S}_{\R}^{2}\bigr)\bigr],
\end{align}
where ${a_\R}={A_{\R3}}/[12q(q+2)(q+4)]$, ${b_\R}=(A_{\R22} -2A_{\R3})/[12q(q+2)]$,
${c_\R}=(A_{\R11}-A_{\R22}+A_{\R3})/({12q})$,
$A_{\R11}=A_{\R1}+A_{\R1,\beta\phi}$, $A_{\R22}=A_{\R2}+A_{\R2,\beta\phi}$,
\begin{align*}
A_{\R1}&= 3\phi^{-1}\bm{1}_{n}^{\top}\bm{F}\bm{Z}_{2d}(\bm{Z}-\bm{Z}_2)\bm{Z}_{2d}\bm{F}\bm{1}_n\\
&\quad +6\phi^{-1}\bm{1}_n^{\top}\bm{F}\bm{Z}_{2d}\bm{Z}_2(\bm{Z}-\bm{Z}_2)_d(\bm{F}-\bm{G})\bm{1}_n\\
&\quad - 6\phi^{-1}\bm{1}_n^{\top}\bm{F}[\bm{Z}_{2}^{(2)}\otimes(\bm{Z}-\bm{Z}_2)](2\bm{G}-\bm{F})\bm{1}_n\\
&\quad-6\phi^{-1}\bm{1}_n^{\top}\bm{H}(\bm{Z}-\bm{Z}_2)_d\bm{Z}_{2d}\bm{1}_n,
\end{align*}
\begin{align*}
A_{\R2}&=-3\phi^{-1}\bm{1}_n^{\top}(\bm{F}-\bm{G})(\bm{Z}-\bm{Z}_2)_d \bm{Z}_2(\bm{Z}-\bm{Z}_2)_d(\bm{F}-\bm{G})\bm{1}_n \\
&\quad -6\phi^{-1}\bm{1}_n^{\top}\bm{F} \bm{Z}_{2d}(\bm{Z}-\bm{Z}_2)(\bm{Z}-\bm{Z}_2)_d(\bm{F}-\bm{G})\bm{1}_n\\
&\quad -6\phi^{-1}\bm{1}_n^{\top}(\bm{F}-\bm{G})[(\bm{Z}-\bm{Z}_2)^{(2)}\otimes \bm{Z}_2](\bm{F}-\bm{G})\bm{1}_n\\
&\quad +3\phi^{-1}\bm{1}_n^{\top}\bm{B}(\bm{Z}-\bm{Z}_2)_{d}^{(2)}\bm{1}_n,
\end{align*}
\begin{align*}
A_{\R3}&= 3\phi^{-1}\bm{1}_n^{\top}(\bm{F}-\bm{G})(\bm{Z}-\bm{Z}_2)_d(\bm{Z}-\bm{Z}_2)(\bm{Z}-\bm{Z}_2)_d(\bm{F}-\bm{G})\bm{1}_n\\
&\quad +2\phi^{-1}\bm{1}_n^{\top}(\bm{F}-\bm{G})(\bm{Z}-\bm{Z}_2)^{(3)}(\bm{F}-\bm{G})\bm{1}_n,
\end{align*}
\[
A_{\R1,\beta\phi}={\frac{6q[d_{(3)}+(2-p+q)d_{(2)}]}{nd_{(2)}^{2}}},\qquad
A_{\R2,\beta\phi}={\frac{3q(q+2)}{nd_{(2)}}}.
\]
The notation $(\cdot)_d$ indicates that the off-diagonal elements of the matrix were set equal to zero.
These formulas are valid when $\phi$ is unknown and estimated from the data. When $\phi$ is known, the terms  
$A_{\R1,\beta\phi}$ and $A_{\R2,\beta\phi}$ are zero.
The factor $[1-({c}_{\R}+{b}_{\R}{S}_\R+{a}_{\R}{S}_{\R}^{2})]$ in \eqref{escorecorr}
is regarded as a Bartlett-type correction factor for the score statistic in such a
way that the null distribution of ${S}_{\R}^{*}$
is better approximated by the reference $\chi^2$ distribution than the
distribution of the uncorrected score statistic. The null distribution of $S_\R^*$
is chi-square with approximation error reduced from order $n^{-1}$ to $n^{-2}$.

In the following, we shall derive an improved gradient statistic
for testing $\mathcal{H}_{0}: \bm{\beta}_{1} = \bm{\beta}_{10}$ in GLMs.
All the results regarding the gradient test in GLMs are new.
The basic idea of transforming the gradient test statistic in such a way that it becomes
better approximated by the reference chi-squared distribution is due to
\cite{VargasFerrariLemonte2013}. The corrected gradient statistic
proposed by these authors is obtained by multiplying the original
gradient statistic by a second-degree polynomial
in the original gradient statistic itself,
producing a modified gradient test statistic whose null
distribution has its asymptotic chi-squared approximation error reduced
from $n^{-1}$ to $n^{-2}$. This idea of improving the
gradient statistic is exactly the same as that to improve the score statistic
\citep{CordeiroFerrari1991}. Thus, improved gradient tests may be based on the
corrected gradient statistic which are expected to
deliver more accurate inferences with samples of typical sizes encountered by applied
practitioners.

The Bartlett-type correction factor for the gradient statistic
derived by \cite{VargasFerrariLemonte2013} is very general
in the sense that it is not tied to a particular parametric
model, and hence needs to be tailored for each
application of interest. The general expression can be very difficult
to particularize for specific regression models because it involves complicated
functions of moments of log-likelihood derivatives up to the fourth order.
As we shall see below, we have been able to apply their results for GLMs;
that is, we derive closed-form expressions for the Bartlett-type correction factor
that defines the corrected gradient statistic in this class of models,
allowing for the computation of this factor with minimal effort.
The Bartlett-corrected gradient statistic is given by
\begin{align} \label{gradcorr}
{S}_{\T}^{*}={S}_{\T}\bigl[1-\bigl({c}_{\T}+{b}_{\T}{S}_\T+{a}_{\T}{S}_{\T}^{2}\bigr)\bigr],
\end{align}
where ${a_\T}={A_{\T3}}/[{12q(q+2)(q+4)}]$, ${b_\T}=(A_{\T22} -2A_{\T3})/[12q(q+2)]$,
${c_\T}=(A_{\T11}-A_{\T22}+A_{\T3})/({12q})$,
$A_{\T11}=A_{\T1}+A_{\T1,\beta\phi}$, $A_{\T22}=A_{\T2}+A_{\T2,\beta\phi}$,
\begin{align*}
A_{\T1}&=12\phi^{-1}\bm{1}_n^{\top}(\bm{F}+\bm{G})[\bm{Z}_d\bm{Z}\bm{Z}_d -\bm{Z}_{2d}\bm{Z}_2\bm{Z}_{2d}+\bm{Z}^{(3)}
-\bm{Z}_{2}^{(3)}](\bm{F}+\bm{G})\bm{1}_n\\
&\quad-6\phi^{-1}\bm{1}_n^{\top}(\bm{F}+2\bm{G})[(\bm{Z}+\bm{Z}_2)\otimes(\bm{Z}^{(2)}-\bm{Z}_{2}^{(2)})
+(\bm{Z}-\bm{Z}_2)_d(\bm{Z}\bm{Z}_d+\bm{Z}_2\bm{Z}_{2d})\\
&\qquad\qquad+2\bm{Z}_{2d}(\bm{Z}\bm{Z}_d-\bm{Z}_2\bm{Z}_{2d})+2\bm{Z_{2}}^{(2)}\otimes(\bm{Z}-\bm{Z}_2)](\bm{F}+\bm{G})\bm{1}_n\\
&\quad+3\phi^{-1}\bm{1}_n^{\top}(\bm{F}+2\bm{G})[2(\bm{Z}-\bm{Z}_2)_d \bm{Z}_2 \bm{Z}_{2d}+2\bm{Z}_{2}^{(2)}\otimes(\bm{Z}-\bm{Z}_2)\\
&\qquad\qquad+\bm{Z}_{2d}(\bm{Z}-\bm{Z}_2)\bm{Z}_{2d}+
\bm{Z}_{2d}(\bm{Z}-\bm{Z}_2)(\bm{Z}-\bm{Z}_2)_{d} ](\bm{F}+2\bm{G})\bm{1}_n\\
&\quad-12\phi^{-1}\bm{1}_n^{\top}\bm{D}(\bm{Z}_{d}^{(2)}-\bm{Z}_{2d}^{(2)})\bm{1}_n+6\phi^{-1}\bm{1}_n^{\top}\bm{T}(\bm{Z}-\bm{Z}_2)_d(\bm{Z}_d
+3\bm{Z}_{2d})\bm{1}_n\\
&\quad-6\phi^{-1}\bm{1}_n^{\top}\bm{E}(\bm{Z}-\bm{Z}_2)_d\bm{Z}_{2d}\bm{1}_n,
\end{align*}
\begin{align*}
A_{\T2}&=-3\phi^{-1}\bm{1}_n^{\top}(\bm{F}+2\bm{G})\biggl[\frac{3}{4}(\bm{Z}-\bm{Z}_2)_d(\bm{Z}-\bm{Z}_2)(\bm{Z}-\bm{Z}_2)_d
+\frac{1}{2}(\bm{Z}-\bm{Z}_2)^{(3)}\\
&\qquad\qquad+\bm{Z}_{2d}(\bm{Z}-\bm{Z}_2)(\bm{Z}-\bm{Z}_2)_d+(\bm{Z}-\bm{Z}_2)_d\bm{Z}_2(\bm{Z}-\bm{Z}_2)_d\\
&\qquad\qquad+2\bm{Z}_2\otimes(\bm{Z}-\bm{Z}_2)^{(2)}\biggr](\bm{F}+2\bm{G})\bm{1}_n\\
&\quad+6\phi^{-1}\bm{1}_n^{\top}(\bm{F}+2\bm{G})[(\bm{Z}-\bm{Z}_2)\otimes(\bm{Z}^{(2)}-\bm{Z}_{2}^{(2)}) \\
&\qquad\qquad+(\bm{Z}-\bm{Z}_2)_d(\bm{Z}\bm{Z}_d-\bm{Z}_2\bm{Z}_{2d})](\bm{F}+\bm{G})\bm{1}_n\\
&\quad-3\phi^{-1}\bm{1}_n^{\top}(2\bm{T}-\bm{E})(\bm{Z}-\bm{Z}_2)_{d}^{(2)}\bm{1}_n,
\end{align*}
\begin{align*}
A_{\T3}&=\phi^{-1}\bm{1}_n^{\top}(\bm{F}+2\bm{G})\biggl[\frac{3}{4}(\bm{Z}-\bm{Z}_2)_d(\bm{Z}-\bm{Z}_2)(\bm{Z}-\bm{Z}_2)_d\\
&\qquad\qquad+\frac{1}{2}(\bm{Z}-\bm{Z}_2)^{(3)}\biggr](\bm{F}+2\bm{G})\bm{1}_n,
\end{align*}
\[
A_{\T1,\beta\phi}={\frac{6q[d_{(3)}+(2-p+q)d_{(2)}]}{nd_{(2)}^{2}}},\qquad
A_{\T2,\beta\phi}={\frac{3q(q+2)}{nd_{(2)}}};
\]
when $\phi$ is known $A_{\T1,\beta\phi}$ and $A_{\T2,\beta\phi}$ are zero.  
The detailed derivation of these expressions is presented in the Supplementary Material.
We basically follow similar algebraic developments of \cite{CordeiroFerrariPaula1993}.
The modified statistic $S_{\T}^*$ has a $\chi_q^2$ distribution up to an error of order
$n^{-2}$ under the null hypothesis.

A brief commentary on the quantities $A_{\T11}=A_{\T1}+A_{\T1,\beta\phi}$,
$A_{\T22}=A_{\T2}+A_{\T2,\beta\phi}$ and $A_{\T3}$
that define the improved gradient statistic is in order.
Comments on the quantities that define the improved LR and score
statistics are given in the corresponding articles in which
they were obtained. Note that $A_{\T1}$,
$A_{\T2}$ and $A_{\T3}$ depend heavily on the
particular model matrix $\bm{X}$ in question. They involve the (possibly unknown)
dispersion parameter and the unknown means. Further, they depend on the mean
link function and its first, second and third derivatives.
They also involve the variance function and its first and second derivatives.
Unfortunately, they are not easy to interpret in generality and provide no
indication as to what structural aspects of the model
contribute significantly to their magnitude.
The quantities $A_{\T1,\beta\phi}$ and $A_{\T2,\beta\phi}$
can be regarded as the contribution yielded by the fact that $\phi$
is considered unknown and has to be estimated from the data.
Notice that $A_{\T1,\beta\phi}$ depends on the model matrix only through
its rank, i.e.~the number of regression parameters ($p$), and it also
involves the number of parameters of interest ($q$) in the
null hypothesis. Additionally,  $A_{\T2,\beta\phi}$ involves the
number of parameters of interest. Therefore, it implies that these quantities can be non-negligible if the
dimension of $\bm{\beta}$ and/or the number of tested parameters in the
null hypothesis are not considerably smaller than the sample size.
Finally, note that $A_{\T1,\beta\phi}$ and $A_{\T2,\beta\phi}$ are exactly the same as the corresponding terms in the
improved score statistic.

Notice that the general expressions which define the improved LR, score and
gradient statistics only involve simple operations
on matrices and vectors, and can be easily implemented
in any mathematical or statistical/econometric programming
environment, such as {\tt R} \citep{R2009}, {\tt Ox}
\citep{Doornik09} and {\tt MAPLE} \citep{Abell-et-al-2003}.
Also, all the unknown parameters in the quantities that define
the improved statistics are replaced by their restricted MLEs.
The improved LR, score and gradient tests that employ \eqref{LRcor},
\eqref{escorecorr} and \eqref{gradcorr}, respectively, as test statistics,
follow from the comparison of
$S^{*}_{\LR}$, $S^{*}_{\R}$ and $S^{*}_{\T}$ with
the critical value obtained as the appropriate $\chi_{q}^{2}$ quantile.

We have that, up to an error of order $n^{-2}$, the null distribution
of the improved statistics
$S^{*}_{\LR}$, $S^{*}_{\R}$ and $S^{*}_{\T}$ is $\chi_q^2$.
Hence, if the sample size is large, all improved tests could be recommended,
since their type I error probabilities do not significantly
deviate from the true nominal level. The natural question is how these
tests perform when the sample size is small or of moderate size,
and which one is the most reliable to test hypotheses
in GLMs. In Section \ref{simulations}, we shall
use Monte Carlo simulation experiments to shed some light on this issue.
In addition to the improved tests, for the sake of comparison 
we also consider  the original LR, Wald, score and gradient
tests in the simulation experiments.

\section{Tests on the parameter $\phi$}\label{phi-tests}

In this section, the problem under consideration is that of testing
a composite null hypothesis $\mathcal{H}_{0}:\phi=\phi_{0}$ against
$\mathcal{H}_{a}:\phi\neq\phi_{0}$, where $\phi_{0}$ is a positive specified value
for $\phi$. Here, $\bm{\beta}$ acts as a vector of nuisance parameters.
The likelihood ratio ($S_{\LR}$), Wald ($S_\W$), Rao score ($S_\R$) and gradient ($S_\T$) statistics
for testing $\mathcal{H}_{0}:\phi=\phi_{0}$ can be expressed, respectively, as
\[
S_{\LR} = 2n[a_{1}(\widehat{\phi}) - a_{1}(\phi_{0})
- (\widehat{\phi} - \phi_{0})a_{1}'(\widehat{\phi})],
\]
\[
S_{\W} = -n(\widehat{\phi} - \phi_{0})^2a_{1}''(\widehat{\phi}),
\]
\[
S_{\R} = -\frac{n[a_{1}'(\widehat{\phi}) - a_{1}'(\phi_{0})]^2}{a_{1}''(\phi_{0})},
\]
\[
S_{\T} = n[a_{1}'(\phi_{0}) - a_{1}'(\widehat{\phi})](\widehat{\phi} - \phi_{0}),
\]
where $\widehat{\phi}$ is the MLE of $\phi$. For example, we have $a_1(\phi) = \log(\phi)/2$ for
the normal and inverse normal models, which yields
\[
S_{\LR} = n\biggl[\log\biggl(\frac{\widehat{\phi}}{\phi_{0}}\biggr)
- \biggl(\frac{\widehat{\phi} - \phi_{0}}{\widehat{\phi}}\biggr)\biggr],
\]
\[
S_{\W} = S_{\R} = \frac{n}{2}\biggl[\frac{\widehat{\phi} - \phi_{0}}{\widehat{\phi}}\biggr]^2,
\qquad
S_{\T} = \frac{n}{2}\frac{(\widehat{\phi} - \phi_{0})^2}{\phi_{0}\widehat{\phi}}.
\]
For the gamma model, we have $a_1(\phi)=\phi\log(\phi) - \log[\Gamma(\phi)]$ and hence
\[
S_{\LR} = 2n\biggl[\phi_{0}\log\biggl(\frac{\widehat{\phi}}{\phi_{0}}\biggr)
-\log\biggl(\frac{\Gamma(\widehat{\phi})}{\Gamma(\phi_{0})}\biggr)
-(\widehat{\phi} - \phi_{0})(1-\psi(\widehat{\phi}))\biggr],
\]
\[
S_{\W} = n[\widehat{\phi}\psi'(\widehat{\phi}) - 1]
\frac{(\widehat{\phi} - \phi_{0})^2}{\widehat{\phi}},
\]
\[
S_{\R} = \frac{n\phi_{0}\{\log(\widehat{\phi}/\phi_{0}) - [\psi(\widehat{\phi}) - \psi(\phi_{0})]\}}
{\phi_{0}\psi'(\phi_{0}) - 1},
\]
\[
S_{\T} = n(\widehat{\phi}-\phi_{0})\biggl[\log\biggl(\frac{\phi_{0}}{\widehat{\phi}}\biggr)
+\psi(\widehat{\phi}) - \psi(\phi_{0})\biggr],
\]
where $\psi'(\cdot)$ denotes the trigamma function.
Under $\mathcal{H}_0$, these statistics have a $\chi_1^2$ distribution
up to an error of order $n^{-1}$.

From \cite{Cordeiro87}, the Bartlett-corrected LR statistic for testing
$\mathcal{H}_{0}:\phi=\phi_{0}$ is given by $S_{\LR}^* = S_{\LR}/[1 + \epsilon(\phi_{0}, p)]$,
where
\[
\epsilon(\phi,p) = -\frac{p(p-2)}{4nd_{(2)}}+\frac{2pd_{(3)}+d_{(4)}}{4nd_{(2)}^2}-\frac{5d_{(3)}^2}{12nd_{(2)}^3},
\]
where $d_{(4)}=d_{(4)}(\phi)=\phi^4a_{1}''''(\phi)$, with
$a_{1}''''(\phi)=\dd a_{1}'''(\phi)/\dd\phi$. Note that $\epsilon(\phi, p)$
depends on the model matrix only through its rank. More specifically, it is
a second degree polynomial in $p$ divided by $n$.
Hence, $\epsilon(\phi, p)$ can be non-negligible if the dimension of
$\bm{\beta}$ is not considerably smaller than the sample size.
It is also noteworthy that $\epsilon(\phi, p)$ depends on $\phi$ but not on
$\bm{\beta}$. The Bartlett-corrected score statistic to test $\mathcal{H}_{0}:\phi=\phi_{0}$
is given by (\ref{escorecorr}),
where ${a_\R}={A_{\R3}}/180$, ${b_\R}=(A_{\R2} -2A_{\R3})/36$,
${c_\R}=(A_{\R1}-A_{\R2}+A_{\R3})/12$, with
\[
A_{\R1} = -\frac{3p(p-2)}{nd_{(2)}^2},\quad
A_{\R2} = -\frac{3(2pd_{(3)} + d_{(4)})}{nd_{(2)}^2},\quad
A_{\R3} = -\frac{5d_{(3)}^2}{nd_{(2)}^3};
\]
see \cite{CordeiroFerrariPaula1993}.
It should be emphasized that these expressions are
quite simple and depend on the model only through
the rank of $\bm{X}$ and $\phi$. They do not involve the unknown
$\bm{\beta}$.  

Next, we shall derive an improved gradient statistic
to test the null hypothesis $\mathcal{H}_{0}:\phi=\phi_{0}$.
After some algebraic manipulations, we define the
improved gradient statistic as
${S}_{\T}^{*}={S}_{\T}\bigl[1-\bigl({c}_{\T}+{b}_{\T}{S}_\T+{a}_{\T}{S}_{\T}^{2}\bigr)\bigr]$,
where ${a_\T}={A_{\T3}}/180$, ${b_\T}=(A_{\T2} - 2A_{\T3})/36$,
${c_\T}=(A_{\T1}-A_{\T2}+A_{\T3})/12$, with
\[
A_{\T1} = -\frac{3p(p+2)}{nd_{(2)}}-\frac{3(3pd_{(3)}-4d_{(4)})}{nd_{(2)}^2}-\frac{18d_{(3)}^2}{nd_{(2)}^3},
\]
\[
A_{\T2} =-\frac{3(pd_{(3)}-d_{(4)})}{nd_{(2)}^2}-\frac{33d_{(3)}^2}{4nd_{(2)}^3},\qquad
A_{\T3} =-\frac{5d_{(3)}^2}{4nd_{(2)}^3}.
\]
Again, the formulas for the $A$'s are very simple, depend on $\bm{X}$ only through its
rank and do not depend on the unknown parameter $\bm{\beta}$. The $A$'s are all
evaluated at $\phi_{0}$. The detailed derivation of the  $A$'s is presented in the Supplementary Material.

Under the null hypothesis, the  adjusted
statistics ${S}_{\LR}^{*}$, ${S}_{\R}^{*}$ and ${S}_{\T}^{*}$
have a $\chi_1^2$ distribution up to an error of order $n^{-2}$.
The improved LR, score and gradient tests follow from the comparison of
$S^{*}_{\LR}$, $S^{*}_{\R}$ and $S^{*}_{\T}$ with
the critical value obtained as the appropriate $\chi_{1}^{2}$ quantile.

\section{Finite-sample power and size properties}\label{simulations}

In what follows, we shall report the results from Monte Carlo simulation experiments
in order to compare the performance of the usual LR ($S_{\LR}$), Wald ($S_{\W}$),
score ($S_{\R}$) and gradient ($S_{\T}$) tests, and the
improved LR ($S_{\LR}^*$), score ($S_{\R}^*$) and gradient ($S_{\T}^*$)
tests in small- and moderate-sized samples for testing hypotheses in the class of GLMs.
We assume that
\[
d(\mu_{l}) = \log(\mu_{l})=\eta_{l} = \beta_{1}x_{l1} + \beta_{2}x_{l2} + \cdots + \beta_{p}x_{lp},
\qquad l=1,\ldots,n,
\]
where $\phi>0$ is assumed unknown and it is the same for all observations.
The number of Monte Carlo replications was 15,000, and the nominal levels
of the tests were $\alpha = 10\%, 5\%$ and $1\%$. The simulations were carried
out using the {\tt Ox} matrix programming language \citep{Doornik09}, which
is freely distributed for academic purposes and available at http://www.doornik.com.
All log-likelihood maximizations with respect to the model parameters were carried out using
the BFGS quasi-Newton method with analytic first derivatives
through {\sf MaxBFGS} subroutine. All the regression parameters,
except those fixed at the null hypothesis, were set equal to one.
The simulation results are based on the gamma and inverse normal regression models.
For the gamma model, we set $\phi=1$ and the covariate values were
selected as random draws from the $\mathcal{U}(0,1)$
distribution. We set $\phi=3$ and selected the covariate values
 as random draws from the $\mathcal{N}(0,1)$
distribution for the inverse normal model.
For each fixed $n$, the covariate values were kept constant
throughout the experiment for both gamma and inverse normal regression models.

We report the null rejection rates
of $\mathcal{H}_0:\beta_{1}=\cdots=\beta_{q}=0$ for all the tests at the $10\%$, $5\%$ and $1\%$
nominal significance levels, i.e.~the percentage of times that the corresponding statistics
exceed the $10\%$, $5\%$ and $1\%$ upper points of the reference $\chi^2$ distribution.
The results are presented in Tables \ref{tab1} and \ref{tab3}
for the gamma model, whereas Tables \ref{tab4} and \ref{tab6} report
the results for the inverse normal model. Entries are percentages.
We consider different values for $p$ (number of regression parameters), $q$ (number
of tested parameters in the null hypothesis) and $n$ (sample size).

The figures in Tables \ref{tab1} to \ref{tab6} reveal important information.
The test that uses the Wald statistic ($S_\W$) is markedly liberal
(over-rejecting the null hypothesis more frequently than expected based on the selected nominal
level), more so as the number of tested parameters in the null
hypothesis ($q$) and the number of regression parameters ($p$)
increase. For example, if $p=4$, $\alpha=10\%$ and
$n=20$, the null rejection rates are 18.04\% (for $q=1$), 21.27\% (for $q=2$)
and 24.13\% (for $q=3$) for the gamma model (Table \ref{tab1}),
whereas we have 18.39\% (for $q=1$), 33.67\% (for $q=2$) and 41.17\% (for $q=3$)
for the inverse normal model (Table \ref{tab4}). Also, if $q=2$, $\alpha=5\%$
and $n=25$, the null rejection rates are 12.11\% (for $p=4$) and 16.45\% (for $p=6$)
for the gamma model (see Tables \ref{tab1} and \ref{tab3}),
and 19.35\% (for $p=4$) and 21.49\% (for $p=6$)
for the inverse normal model (see Tables \ref{tab4} and \ref{tab6}).
Notice that  the test which uses the original LR statistic ($S_{\LR}$)
is also liberal, but less size distorted than the Wald
test. In the above examples, the null rejection rates are 14.31\% (for $q=1$),
15.25\% (for $q=2$) and 15.78\% (for $q=3$)
for the gamma model (Table \ref{tab1}), and 14.99\% (for $q=1$),
18.76\% (for $q=2$) and 20.03\% (for $q=3$)
for the inverse normal model (Table \ref{tab4}). Also, we have 7.89\% (for $p=4$) and 9.89\% (for $p=6$)
for the gamma model (see Tables \ref{tab1} and \ref{tab3}),
and 9.00\% (for $p=4$) and 11.99\% (for $p=6$)
for the inverse normal model (see Tables \ref{tab4} and \ref{tab6}).
The original score ($S_\R$) and gradient ($S_\T$) tests are
also liberal in most of the cases, but less size distorted than the original
LR and Wald tests in all cases. It is noticeable that the original score test is much
less liberal than the original LR and Wald tests and
slightly less liberal than the original gradient test.

As pointed out above, the usual score and gradient tests
are less size distorted than the original LR and Wald tests.
However, their null rejection rates can also deviate considerably
of the significance levels of the test.
For example, if $p=6$, $q=2$, $\alpha=10\%$ and $n=20$,
the null rejection rates are 11.20\% ($S_\R$) and 15.24\% ($S_\T$)
for the gamma model (see Table \ref{tab3}),
and 12.21\% ($S_\R$) and 14.85\% ($S_\T$)
for the inverse normal model (see Table \ref{tab6}).
On the other hand, the improved LR, score and
gradient tests that employ ${S}^{*}_{\LR}$, ${S}^{*}_{\R}$ and ${S}^{*}_{\T}$
as test statistics, respectively, are less size distorted
than the usual LR, Wald, score and gradient tests
for testing hypotheses in GLMs; that is,
the impact of the number of regressors and the number of tested
parameters in the null hypothesis are much less important for the improved
tests. Among the improved tests, the test that uses the statistic ${S}^{*}_{\LR}$
presents the worst performance, displaying null rejection rates
more size distorted than the improved score and gradient tests
in most of the cases. For example, if $p=6$, $q=4$, $\alpha=10\%$ and $n=20$,
the null rejection rates of ${S}^{*}_{\LR}$, ${S}^{*}_{\R}$ and ${S}^{*}_{\T}$
are, 12.15\%, 10.20\% and 10.03\%, respectively,
for the gamma model (see Table \ref{tab3}), and  14.21\%, 10.81\% and 10.60\%, respectively,
for the inverse normal model (see Table \ref{tab6}).
The improved score and gradient tests produced null rejection rates that are very
close to the nominal levels in all the cases considered.
Finally, the figures in Tables \ref{tab1} to \ref{tab6}
show that the null rejection rates of all tests approach the corresponding nominal
levels as the sample size grows, as expected.
\begin{table}[!htp]
\centering
{\footnotesize
\caption{Null rejection rates (\%) for $\mathcal{H}_0:\beta_{1}=\cdots=\beta_{q}=0$ with $p=4$;
gamma model.}    \label{tab1}
\begin{tabular}{lllrrrrrrr}  \hline
$q$&    $n$     & $\alpha(\%)$ & ${S}_\W$ & ${S}_{\LR}$ & ${S}_{\R}$ & ${S}_{\T}$ & ${S}^{*}_{\LR}$ & ${S}^{*}_{\R}$ & ${S}^{*}_{\T}$ \\   \hline
3&   20    & 10    & 24.13 & 15.78 & 9.23  & 10.18 & 10.73 & 9.93  & 9.73 \\
 &         & 5     & 16.73 & 8.82  & 4.10   & 4.37  & 5.54  & 4.97  & 4.77 \\
 &         & 1     & 7.59  & 2.43  & 0.55  & 0.37  & 1.13  & 0.99  & 0.73 \\
 &         &       &       &       &       &       &       &       &  \\
 &   25    & 10    & 21.17 & 14.23 & 9.32  & 10.26 & 10.45 & 9.91  & 9.91 \\
 &         & 5     & 13.92 & 7.82  & 4.16  & 4.39  & 5.19  & 4.96  & 4.69 \\
 &         & 1     & 5.53  & 1.81  & 0.63  & 0.46  & 0.99  & 0.90   & 0.76 \\
 &         &       &       &       &       &       &       &       &  \\
 &   30    & 10    & 19.81 & 13.97 & 10.24 & 10.55 & 10.70 & 10.64 & 10.28 \\
 &         & 5     & 12.68 & 7.62  & 4.79  & 5.08  & 5.48  & 5.37  & 5.25 \\
 &         & 1     & 4.75  & 1.87  & 0.84  & 0.67  & 1.17  & 1.10   & 0.97 \\\hline
2&    20    & 10    & 21.27 & 15.25 & 9.41  & 11.71 & 10.68 & 9.96  & 10.07 \\
 &          & 5     & 14.03 & 8.71  & 4.03  & 5.61  & 5.61  & 4.99  & 5.01 \\
 &          & 1     & 6.15  & 2.36  & 0.39  & 0.73  & 1.17  & 0.93  & 0.91 \\
 &          &       &       &       &       &       &       &       &  \\
 &    25    & 10    & 19.05 & 14.3  & 10.08 & 11.69 & 10.59 & 10.27 & 10.07 \\
 &          & 5     & 12.11 & 7.89  & 4.69  & 5.59  & 5.41  & 5.23  & 5.07 \\
 &          & 1     & 4.51  & 1.91  & 0.63  & 0.63  & 0.95  & 0.90   & 0.76 \\
 &          &       &       &       &       &       &       &       &  \\
 &    30    & 10    & 17.52 & 13.56 & 10.10  & 11.46 & 10.72 & 10.46 & 10.31 \\
 &          & 5     & 10.70 & 7.34  & 4.45  & 5.30   & 5.15  & 4.96  & 4.94 \\
 &          & 1     & 3.79  & 1.83  & 0.65  & 0.76  & 1.05  & 0.99  & 0.90 \\ \hline
1&    20    & 10    & 18.04 & 14.31 & 10.62 & 13.02 & 10.97 & 10.29 & 10.39 \\
 &          & 5     & 11.55 & 8.24  & 5.15  & 6.68  & 5.68  & 5.19  & 5.23 \\
 &          & 1     & 4.47  & 2.28  & 0.76  & 1.15  & 1.17  & 0.94  & 0.96 \\
 &          &       &       &       &       &       &       &       &  \\
 &    25    & 10    & 16.52 & 13.58 & 10.30  & 12.38 & 10.67 & 10.49 & 10.39 \\
 &          & 5     & 10.16 & 7.45  & 4.81  & 6.35  & 5.47  & 5.30   & 5.20 \\
 &          & 1     & 3.67  & 1.91  & 0.71  & 1.19  & 1.19  & 1.05  & 1.11 \\
 &          &       &       &       &       &       &       &       &  \\
 &    30    & 10    & 15.39 & 12.39 & 9.73  & 11.61 & 10.15 & 10.01 & 9.89 \\
 &          & 5     & 9.22  & 6.64  & 4.69  & 5.97  & 5.19  & 5.09  & 5.04 \\
 &          & 1     & 2.95  & 1.45  & 0.74  & 0.99  & 0.99  & 0.97  & 0.95 \\ \hline
\end{tabular}      }
\end{table}
\begin{table}[!htp]
\centering
{\footnotesize
\caption{Null rejection rates (\%) for $\mathcal{H}_0:\beta_{1}=\cdots=\beta_{q}=0$ with $p=6$;
gamma model.}    \label{tab3}
\begin{tabular}{lllrrrrrrr}  \hline
$q$&    $n$     & $\alpha(\%)$ & ${S}_\W$ & ${S}_{\LR}$ & ${S}_{\R}$ & ${S}_{\T}$ & ${S}^{*}_{\LR}$ & ${S}^{*}_{\R}$ & ${S}^{*}_{\T}$ \\   \hline
4&    20    & 10    & 35.69& 20.23  & 9.77  & 11.57 & 12.15 & 10.20  & 10.03 \\
 &          & 5     & 27.14& 12.40   & 4.45  & 4.99  & 6.61  & 4.97  & 4.82 \\
 &          & 1     & 15.31& 3.99   & 0.64  & 0.37  & 1.47  & 0.76  & 0.69 \\
 &          &       &       &       &       &       &       &       &  \\
 &    25    & 10    & 28.99& 17.23  & 8.96  & 11.04 & 11.01 & 10.09 & 9.82 \\
 &          & 5     & 20.45& 9.98   & 3.95  & 4.57  & 5.59  & 4.81  & 4.51 \\
 &          & 1     & 9.83 & 2.77   & 0.55  & 0.45  & 1.15  & 0.91  & 0.71 \\
 &          &       &       &       &       &       &       &       &  \\
 &    30    & 10    & 25.41& 15.97  & 8.97  & 10.65 & 10.65 & 9.93  & 9.73 \\
 &          & 5     & 17.52& 8.93   & 3.83  & 4.65  & 5.29  & 4.81  & 4.69 \\
 &          & 1     & 8.05 & 2.39   & 0.54  & 0.59  & 1.13  & 0.25  & 0.82 \\ \hline
3&    20    & 10    & 31.15& 19.87  & 10.13 & 13.57 & 11.95 & 10.33 & 10.14 \\
 &          & 5     & 22.71& 12.17  & 4.60   & 6.18  & 6.17  & 5.47  & 4.92 \\
 &          & 1     & 11.53& 3.79   & 0.51  & 0.61  & 1.23  & 0.89  & 0.70 \\
 &          &       &       &       &       &       &       &       &  \\
 &    25    & 10    & 27.51& 17.65  & 9.75  & 12.51 & 10.95 & 10.03 & 9.95 \\
 &          & 5     & 19.09& 9.82   & 4.27  & 5.75  & 5.63  & 4.95  & 4.77 \\
 &          & 1     & 8.61& 2.66    & 0.59  & 0.76  & 1.16  & 0.83  & 0.81 \\
 &          &       &       &       &       &       &       &       &  \\
 &    30    & 10    & 22.89& 15.43  & 9.43  & 11.87 & 10.34 & 9.55  & 9.65 \\
 &          & 5     & 15.51& 8.56   & 4.15  & 5.69  & 5.32  & 4.62  & 4.70 \\
 &          & 1     & 6.45& 2.15    & 0.51  & 0.67  & 0.90   & 0.74  & 0.69 \\  \hline
2&    20    & 10    & 26.77& 18.80   & 11.20  & 15.24 & 11.8  & 10.58 & 10.27 \\
 &          & 5     & 18.88& 11.39  & 5.11  & 7.57  & 6.17  & 5.45  & 4.88 \\
 &          & 1     & 8.72 & 3.43   & 0.62  & 1.10  & 1.32  & 1.01  & 0.85 \\
 &          &       &       &       &       &       &       &       &  \\
 &    25    & 10    & 24.36& 16.92  & 10.56 & 14.03 & 11.28 & 10.29 & 10.4 \\
 &          & 5     & 16.45& 9.89   & 4.83  & 7.03  & 5.67  & 5.02  & 4.97 \\
 &          & 1     & 7.07 & 2.70    & 0.71  & 1.04  & 1.05  & 1.03  & 0.84 \\
 &          &       &       &       &       &       &       &       &  \\
 &    30    & 10    & 20.61& 15.05  & 10.36 & 12.99 & 10.71 & 10.08 & 10.01 \\
 &          & 5     & 13.5 & 8.51   & 4.70  & 6.37  & 5.48  & 4.86  & 5.04 \\
 &          & 1     & 5.28& 2.29    & 0.61  & 0.97  & 1.00     & 0.81  & 0.82 \\ \hline
\end{tabular}  }
\end{table}
\begin{table}[!htp]
\centering
{\footnotesize
\caption{Null rejection rates (\%) for $\mathcal{H}_0:\beta_{1}=\cdots=\beta_{q}=0$ with $p=4$;
inverse normal model.}    \label{tab4}
\begin{tabular}{lllrrrrrrr}  \hline
$q$&    $n$     & $\alpha(\%)$ & ${S}_\W$ & ${S}_{\LR}$ & ${S}_{\R}$ & ${S}_{\T}$ & ${S}^{*}_{\LR}$ & ${S}^{*}_{\R}$ & ${S}^{*}_{\T}$ \\   \hline
3&    20    & 10    & 41.17 & 20.03 & 8.85  & 7.68  & 12.95 & 10.44 & 10.01 \\
 &          & 5     & 32.72 & 11.96 & 4.64  & 2.81  & 6.84  & 2.85  & 4.99 \\
 &          & 1     & 20.69 & 3.55  & 1.19  & 0.26  & 1.55  & 0.02  & 0.99 \\
 &          &       &       &       &       &       &       &       &  \\
 &    25    & 10    & 37.37 & 17.61 & 8.67  & 7.83  & 11.85 & 10.11 & 9.50 \\
 &          & 5     & 29.59 & 10.17 & 4.44  & 2.83  & 5.95  & 4.10   & 4.44 \\
 &          & 1     & 18.4  & 2.82  & 0.97  & 0.23  & 1.35  & 0.15  & 0.77 \\
 &          &       &       &       &       &       &       &       &  \\
 &    30    & 10    & 34.66 & 15.86 & 8.76  & 7.98  & 10.95 & 9.86  & 9.05 \\
 &          & 5     & 26.33 & 8.69  & 4.40   & 2.91  & 5.43  & 4.45  & 4.21 \\
 &          & 1     & 14.88 & 2.26  & 0.85  & 0.17  & 1.17  & 0.41  & 0.65 \\ \hline
2&    20    & 10    & 33.67 & 18.76 & 7.81  & 9.15  & 12.51 & 10.93 & 9.90 \\
 &          & 5     & 25.66 & 11.19 & 3.24  & 3.27  & 6.53  & 5.29  & 4.71 \\
 &          & 1     & 15.01 & 3.29  & 0.39  & 0.21  & 1.39  & 0.93  & 0.73 \\
 &          &       &       &       &       &       &       &       &  \\
 &    25    & 10    & 27.04 & 16.03 & 9.62  & 10.51 & 11.32 & 10.22 & 9.95 \\
 &          & 5     & 19.35 & 9.00     & 4.47  & 4.55  & 5.86  & 4.74  & 4.93 \\
 &          & 1     & 9.99  & 2.42  & 0.90  & 0.31  & 1.21  & 0.79  & 0.71 \\
 &          &       &       &       &       &       &       &       &  \\
 &    30    & 10    & 23.68 & 15.14 & 8.96  & 11.03 & 11.23 & 10.52 & 10.46 \\
 &          & 5     & 16.35 & 8.77  & 4.09  & 4.76  & 5.71  & 5.09  & 4.92 \\
 &          & 1     & 7.35  & 2.20   & 0.63  & 0.49  & 1.30   & 0.99  & 0.85 \\ \hline
1&    20    & 10    & 18.39 & 14.99 & 11.46 & 13.19 & 10.83 & 10.12 & 9.94 \\
 &          & 5     & 11.95 & 8.48  & 5.67  & 6.46  & 5.49  & 5.09  & 4.97 \\
 &          & 1     & 4.89  & 2.49  & 1.06  & 1.11  & 1.26  & 1.07  & 0.97 \\
 &          &       &       &       &       &       &       &       &  \\
 &    25    & 10    & 20.06 & 14.83 & 9.50   & 12.41 & 11.57 & 10.73 & 10.63 \\
 &          & 5     & 13.27 & 8.46  & 4.34  & 5.86  & 5.68  & 5.53  & 5.19 \\
 &          & 1     & 5.31  & 2.17  & 0.57  & 0.77  & 1.19  & 1.19  & 0.96 \\
 &          &       &       &       &       &       &       &       &  \\
 &    30    & 10    & 15.41 & 13.19 & 10.87 & 11.87 & 10.32 & 10.12 & 9.99 \\
 &          & 5     & 9.48  & 7.39  & 5.19  & 5.83  & 5.34  & 4.95  & 5.14 \\
 &          & 1     & 3.17  & 1.69  & 0.75  & 0.85  & 0.98  & 0.81  & 0.84 \\  \hline
\end{tabular} }
\end{table}
\begin{table}[!htp]
\centering
{\footnotesize
\caption{Null rejection rates (\%) for $\mathcal{H}_0:\beta_{1}=\cdots=\beta_{q}=0$ with $p=6$;
inverse normal model.}    \label{tab6}
\begin{tabular}{lllrrrrrrr}  \hline
$q$&    $n$     & $\alpha(\%)$ & ${S}_\W$ & ${S}_{\LR}$ & ${S}_{\R}$ & ${S}_{\T}$ & ${S}^{*}_{\LR}$ & ${S}^{*}_{\R}$ & ${S}^{*}_{\T}$ \\   \hline
4&    20    & 10    & 51.10  & 26.65 & 7.64  & 8.69  & 14.21 & 10.81 & 10.60\\
 &          & 5     & 42.37 & 16.91 & 3.00     & 2.79  & 7.45  & 5.13  & 4.96 \\
 &          & 1     & 27.45 & 5.77  & 0.39  & 0.25  & 1.70   & 0.94  & 0.97 \\
 &          &       &       &       &       &       &       &       &  \\
 &    25    & 10    & 45.01 & 22.11 & 10.12 & 9.67  & 12.75 & 10.64 & 10.07 \\
 &          & 5     & 36.47 & 13.41 & 5.49  & 4.13  & 6.51  & 4.68  & 5.31 \\
 &          & 1     & 23.18 & 4.02  & 1.37  & 0.40   & 1.55  & 0.19  & 1.04 \\
 &          &       &       &       &       &       &       &       &  \\
 &    30    & 10    & 49.01 & 22.01 & 8.86  & 8.45  & 12.75 & 10.63 & 9.58 \\
 &          & 5     & 39.98 & 13.09 & 4.32  & 3.10   & 6.46  & 4.90   & 4.51 \\
 &          & 1     & 26.23 & 3.79  & 0.83  & 0.23  & 1.44  & 0.49  & 0.80 \\  \hline
3&    20    & 10    & 43.03 & 24.72 & 11.5  & 12.83 & 14.21 & 11.52 & 11.00 \\
 &          & 5     & 34.61 & 15.87 & 5.85  & 5.47  & 7.92  & 5.69  & 5.41 \\
 &          & 1     & 21.27 & 5.65  & 1.29  & 0.51  & 1.86  & 0.63  & 1.03 \\
 &          &       &       &       &       &       &       &       &  \\
 &    25    & 10    & 35.32 & 21.32 & 10.25 & 12.55 & 12.91 & 10.83 & 10.84 \\
 &          & 5     & 26.80 & 12.97 & 4.87  & 5.40   & 6.86  & 5.39  & 5.27 \\
 &          & 1     & 14.36 & 3.95  & 0.94  & 0.71  & 1.59  & 1.09  & 1.09 \\
 &          &       &       &       &       &       &       &       &  \\
 &    30    & 10    & 32.05 & 18.35 & 9.99  & 10.67 & 11.35 & 10.39 & 9.78 \\
 &          & 5     & 23.43 & 10.53 & 4.45  & 4.30   & 5.86  & 4.87  & 4.67 \\
 &          & 1     & 12.32 & 2.95  & 0.72  & 0.33  & 1.16  & 0.87  & 0.71 \\  \hline
2&    20    & 10    & 32.61 & 21.55 & 12.21 & 14.85 & 13.33 & 11.1  & 10.59 \\
 &          & 5     & 24.12 & 13.76 & 6.02  & 7.44  & 7.41  & 5.47  & 5.18 \\
 &          & 1     & 13.15 & 4.70   & 0.93  & 0.85  & 1.71  & 0.95  & 0.83 \\
 &          &       &       &       &       &       &       &       &  \\
 &    25    & 10    & 29.61 & 19.30  & 11.42 & 14.28 & 12.42 & 10.51 & 10.54 \\
 &          & 5     & 21.49 & 11.99 & 5.47  & 7.09  & 6.78  & 5.29  & 5.58 \\
 &          & 1     & 10.61 & 3.76  & 0.91  & 0.91  & 1.50  & 1.03  & 0.94 \\
 &          &       &       &       &       &       &       &       &  \\
 &    30    & 10    & 28.01 & 17.51 & 9.33  & 11.83 & 11.29 & 10.27 & 10.07 \\
 &          & 5     & 20.23 & 10.08 & 4.39  & 5.20   & 5.59  & 5.02  & 4.83 \\
 &          & 1     & 10.01 & 2.64  & 0.90   & 0.75  & 1.11  & 0.98  & 1.05 \\ \hline
\end{tabular} }
\end{table}%

Tables \ref{tab7} and \ref{tab8} contain the nonnull rejection rates (powers) of the tests,
for the gamma and inverse normal regression models, respectively.
Here, $p = 4$, $q=2$, $\alpha = 5\%$ and $n=30$. Data generation was performed
under the alternative hypothesis $\mathcal{H}_a:\beta_{1} = \beta_{2} = \delta$, with different
values of $\delta$. We have only considered the three corrected
tests that use $S_{\LR}^*$, $S_\R^*$ and $S_\T^*$, since the original
LR, Wald, score and gradient tests are considerably size distorted, as noted earlier.
Note that the three improved tests have similar powers in both regression
models. For instance, when $\delta = 0.5$ in the gamma
model (Table \ref{tab7}), the nonnull rejection
rates are 14.57\% ($S_{\LR}^*$), 14.09\% ($S_\R^*$) and 14.05\% ($S_\T^*$).
Additionally, in the inverse normal model with $\delta=-0.20$ (Table \ref{tab8}),
we have 47.63\% ($S_{\LR}^*$), 49.35\% ($S_\R^*$) and
48.94\% ($S_\T^*$). We also note that the powers of the
improved tests increase with $|\delta|$ for both regression models, as expected.
\begin{table}[!htp]
\centering
{\footnotesize
\caption{Nonnull rejection rates (\%): $\alpha=5\%$, $p=4$, $q=2$, $n=30$; gamma model.}\label{tab7}
\begin{tabular}{lrrrrrrrrrr}\hline
$\delta$ &$-4.0$&$-3.0$&$-2.0$&$-1.0$&$-0.50$&0.50&1.0&2.0&3.0&4.0 \\\hline
${S}_{\LR}^{*} $&99.99       &         99.84        &           96.99        &         44.38       &             14.03       &      14.57     &         37.01     &     94.75     &        99.96     &   100.00 \\
${S}_{\R}^{*}$ &99.63        &         99.31        &           92.28        &         40.17       &             12.14       &      14.09     &         37.29     &     90.44     &        99.36     &   99.99 \\
${S}_{\T}^{*}$ &99.99        &         99.83        &           96.73        &         43.02       &             13.13       &      14.05     &         36.73     &     94.64     &        99.97     & 100.00\\\hline
\end{tabular} }
\end{table}
\begin{table}[!htp]
\centering
{\footnotesize
\caption{Nonnull rejection rates (\%): $\alpha=5\%$, $p=4$, $q=2$, $n=30$; inverse normal model.}\label{tab8}
\begin{tabular}{lrrrrrrrrrr}\hline
$\delta   $     &  $-0.50$      &   $-0.40$   &   $-0.30$  &      $-0.20$ &     $-0.10$ &     0.10   &    0.20 &  0.30       &   0.40  &   0.50         \\  \hline
$   {S}_{\LR}^*$ &   99.08  &       95.46  &   80.67  &      47.63 &     15.96 &     19.57  &   58.06 &      88.23  &  97.95  &      99.67     \\
$   {S}_{\R}^*$  &  94.54      &    92.17  &   79.29  &     49.35  &   17.65   &    22.74   &  62.48  &      87.91  &  95.03  &      96.88      \\
${S}_{\T}^*$     &  98.94     &    95.79  &   81.64  &      48.94 &     15.90 &     20.55  &  61.19  &      90.36  &  97.91  &    94.82        \\ \hline
\end{tabular}      }
\end{table}

The main findings from the simulation results of this section can be summarized as follows. 
The usual LR, Wald, score and gradient tests can be considerably oversized (liberal) 
to test hypotheses on the model parameters in GLMs,
over-rejecting the null hypothesis much more frequently than expected
based on the selected nominal level. The usual score and gradient
tests can also be considerably undersized (conservative) in some cases,
under-rejecting the null hypothesis much less frequently than expected
based on the selected nominal level. The improved LR, score and gradient tests
tend to overcome these problems, producing null rejection rates which are
close to the nominal levels. Overall, in small to moderate-sized samples, the best performing
tests are the improved score and gradient tests.
These improved tests perform very well
and hence should be recommended to test hypotheses in GLMs.
Additionally, the Wald test should not be recommended to test hypotheses
in this class of models when the sample size is not large, since it is much more liberal 
than the other tests.

\section{Real data illustration}\label{application}

In this section, we shall illustrate an application of
the usual LR, Wald, score and gradient statistics, and the
improved LR, score and gradient statistics for testing hypotheses
in the class of GLMs in a real data set. The computer code for computing
these statistics using the {\sf Ox} matrix programming language
is presented in the Supplementary Material.
We consider the data reported by \cite{freund83} which 
correspond to an experiment to study the size of squid
eaten by sharks and tuna. The study involved measurements
taken on $n=22$ squids. The variables considered in the study are:
weight ($y$) in pounds, rostral length ($x_2$),
wing length ($x_3$), rostral to notch length ($x_4$),
notch to wing length ($x_5$), and width ($x_6$) in inches.
Notice that the regressor variables are characteristics of
the beak, mouth or wing of the squids.

\begin{figure}[!htp]
\begin{center}
\includegraphics[scale=0.55]{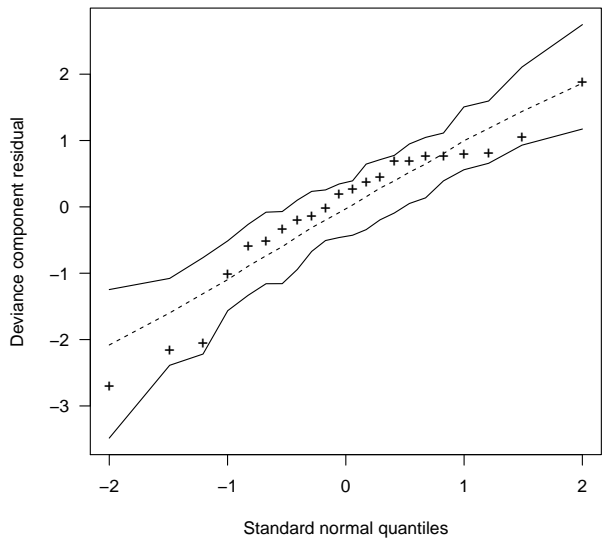}
\caption{Normal probability plot with envelope.}\label{envelopeSQUID}
\end{center}
\end{figure}

We consider the systematic component
\begin{equation}\label{model1}
\log(\mu_{l})=\beta_1 x_{1l} + \beta_2 x_{2l}+ \beta_3 x_{3l}+ \beta_4 x_{4l}+ \beta_5 x_{5l} + \beta_{6}x_{6l},
\qquad l=1,\ldots,22,
\end{equation}
where $x_{1l}=1$ and $\phi>0$ is assumed unknown and it is the same for all observations.
We assume a gamma distribution for the response variable $y$ (weight),
that is, $y_l\sim\mbox{Gamma}(\mu_{l},\phi)$, for $l=1,\ldots,22$.
Figure \ref{envelopeSQUID} presents the normal probability plot
with generated envelopes for the deviance component residual
\citep[see, for example,][]{McCullaghNelder1989} of the
regression model \eqref{model1} fitted to the data. It reveals
that the assumption of the gamma distribution for the data
seems suitable, since there are no observations falling
outside the envelope. The MLEs of the regression
parameters (asymptotic standard errors in parentheses) are:
$\widehat{\beta}_{1} = -2.2899\,(0.2001)$, $\widehat{\beta}_{2} = 0.4027\,(0.5515)$,
$\widehat{\beta}_{3} = -0.4362\,(0.5944)$, $\widehat{\beta}_{4} = 1.2916\,(1.3603)$,
$\widehat{\beta}_{5} = 1.9420\,(0.7844)$ and $\widehat{\beta}_{6} = 2.1394\,(1.0407)$.
The MLE of the precision parameter is $\widehat{\phi}=44.001\,(13.217)$.

Suppose the interest lies in testing the null hypothesis $\mathcal{H}_0:\beta_{4}=\beta_{5}=0$
against a two-sided alternative hypothesis; that is, we want to verify whether there is a significant joint effect
of rostral to notch length and notch to wing length on the mean weight of squids. 
For testing $\mathcal{H}_0$, the observed values of $S_\W$, $S_{\LR}$, $S_{\R}$, $S_{\T}$, $S_{\LR}^*$,
$S_{\R}^*$ and $S_{\T}^*$, and the corresponding p-values are listed in Table \ref{tab10}.
It is noteworthy that one rejects the null hypothesis at the 10\% nominal level
when the inference is based on the usual LR, Wald, score and gradient tests,
and on the improved LR test. However, a different decision is reached when the improved
score and gradient tests are employed. Recall from the previous section that the unmodified tests
are size distorted when the sample is small (here, $n=22$) and are considerably affected by
the number of tested parameters in the null hypothesis (here, $q=2$)
and by the number of regression parameters in the model
(here, $p=6$), which leads us to mistrust the
inference delivered by these tests. Moreover, recall from
our simulation results that the improved LR test can also be oversized,
and hence the improved score and gradient tests should be preferable.
Therefore, on the basis of the adjusted score and gradient tests, the null
hypothesis $\mathcal{H}_0:\beta_{4}=\beta_{5}=0$ should not be
rejected at the 10\% nominal level. Notice that the test
which uses the statistic $S_{\W}$ rejects the
null hypothesis $\mathcal{H}_{0}:\beta_4=\beta_{5}=0$
even at the 5\% nominal level. It confirms the liberal behavior
of the Wald test in testing hypotheses in GLMs, as evidenced
by our Monte Carlo experiments.
\begin{table}[!htp]
\centering
{\footnotesize
\caption{Statistics and p-values: $\mathcal{H}_{0}:\beta_4=\beta_{5}=0$.}   \label{tab10}
\begin{tabular}{lrrrrrrr} \hline
  &  $S_\W$ & $S_{\LR}$ & $S_{\R}$ & $S_{\T}$ & $S_{\LR}^*$ & $S_{\R}^*$ & $S_{\T}^*$\\\hline
Observed value     &7.0659  &5.8976  &4.8382   &5.1193    &4.6380   &4.0842  &4.3239    \\
p-value            &0.0292  &0.0524  &0.0890   &0.0773    &0.0984   &0.1298  &0.1151    \\  \hline
\end{tabular} }
\end{table}

Let $\mathcal{H}_{0}:\beta_{2} = \beta_{3}=\beta_{4} = 0$ be now
the null hypothesis of interest, that is, the exclusion of the  covariates
rostral length, wing length and rostral to notch length
from the regression model \eqref{model1}.
The null hypothesis is not rejected at the 10\% nominal level
by all the tests, but we note that the corrected score and gradient
tests yield larger p-values.
The test statistics are $S_{\LR}=1.3272$, $S_\W=1.4297$,
$S_{\R}=1.2321$, $S_{\T}=1.2876$, $S_{\LR}^*=1.0625$,
$S_{\R}^*=0.9649$ and $S_{\T}^*=1.0044$,
the corresponding p-values being 0.7227, 0.6986,
0.7453, 0.7321, 0.7861, 0.8097 and  0.8002.
We proceed by removing $x_2$, $x_3$ and $x_4$
from model \eqref{model1}. We then estimate
\[
\log(\mu_{l})=\beta_1 x_{1l} + \beta_5 x_{5l} + \beta_{6}x_{6l},\qquad l=1,\ldots,22.
\]
The parameter estimates are (asymptotic standard errors in parentheses):
$\widehat{\beta}_{1} = -2.1339\,(0.1358)$, $\widehat{\beta}_{5} = 2.1428\,(0.3865)$,
$\widehat{\beta}_{6} = 2.9749\,(0.5888)$ and $\widehat{\phi}=41.4440\,(12.446)$.
The null hypotheses  $\mathcal{H}_{0}: \beta_{j} = 0$ ($j=5,6$)
and $\mathcal{H}_{0}: \beta_{5}=\beta_{6} = 0$ are strongly rejected
by the seven tests (unmodified and modified) at the usual significance levels.
Hence, the estimated model is
\[
\widehat{\mu}_l=\e^{-2.1339 + 2.1428x_{5l} + 2.9749x_{6l}},\qquad l=1,\ldots,22.
\]

\section{Discussion}\label{conclusion}

The class of generalized linear models (GLMs) was introduced in 1972 by \cite{NelderWedderburn1972}
as a general framework for handling a range of common statistical models for normal and
non-normal data. This class of models provides a unified approach to many of the most
common statistical procedures used in applied statistics. Many statistical packages
now include facilities for fitting GLMs. In this class of models,
large-samples tests, such as the likelihood ratio (LR), Wald and score tests,  are
the most commonly used statistical tests for testing a composite null hypotheses
on the model parameters, since exact tests are not always available.
An alternative test uses the gradient statistic proposed by
\cite{Terrell2002}. Recently, the gradient test has been the subject
of some research papers. In particular, \cite{Lemonte2011, Lemonte2012} provides
comparison among the local power of the classic tests and the
gradient test in some specific regression models. The author showed
that the gradient test can be an interesting alternative to the classic
tests.

It is well-known that, up to an error of order $n^{-1}$ and under the null
hypothesis, the LR, Wald, score and gradient statistics have a $\chi^2$ distribution
for testing hypotheses concerning the parameters in the class of GLMs.
However, for small sample sizes, the $\chi^2$ distribution may be a poor
approximation to the exact null distribution of these statistics.
In order to overcome this problem, one can use higher order asymptotic theory.
More specifically, one can derive Bartlett and Bartlett-type correction factors
to improve the approximation of the exact null distribution of these statistics
by the $\chi^2$ distribution. The first step in order to improve the
likelihood-based inference in GLMs was provided by \cite{Cordeiro83, Cordeiro87},
who derived Bartlett correction factors for the LR statistic. Next,
\cite{CordeiroFerrariPaula1993} and \cite{CribariFerrari1995} derived
Bartlett-type correction factors for the score statistic.

In this paper, in addition to the improved test statistics
above mentioned, we proposed a Bartlett-corrected gradient statistic
to test composite null hypotheses in GLMs.
To this end, we started from the general results of \cite{VargasFerrariLemonte2013}
and derived Bartlett-type correction factors
for the gradient statistic, which was recently proposed in the statistical
literature. Further, we numerically compared the behavior of  the
original gradient test ($S_{\T}$) and its Bartlett-corrected version ($S_{\T}^*$),
with the Wald test ($S_{\W}$), the LR test ($S_{\LR}$),
the score test ($S_{\R}$) and the Bartlett-corrected LR ($S_{\LR}^*$)
and score ($S_{\R}^*$) tests. We also presented an empirical application to illustrate
the practical usefulness of all test statistics. We show that the finite sample
adjustments can lead to inferences that are different
from those reached based on first order asymptotics.

Our simulation results clearly indicate that
the original LR and Wald tests can be considerably oversized (liberal) and should not
be recommended to test hypotheses in GLMs when the sample is small or of moderate size. The original score and
gradient tests are less size distorted than the original LR and Wald tests,
however, as the number of regression parameters and/or
the number of tested parameters increase, these tests can be
considerably size distorted. Also, the simulations have convincingly shown
that inference based on the modified test statistics can be much more
accurate than that based on the unmodified test statistics.
Overall, our numerical results favor the tests obtained from applying a Bartlett-type
correction to the score and gradient test statistics.
Therefore, we recommend the corrected score and gradient
tests for practical applications. The latter was proposed in the present article.

Finally, it should be emphasized that there has been some effort to produce computer 
codes to compute Bartlett correction factors. For example, \citet{daSilvaCordeiro2009} 
present an {\tt R} program source \citep{R2009} for calculating Bartlett corrections to
improve likelihood ratio tests. We hope to provide a package in {\tt R} 
to compute all the corrected tests, including the 
Bartlett-corrected gradient test derived in this paper.
This is an open problem and we hope to address this issue in a future research.
The advantage of considering the {\tt R} program in relation to others is because
it is a free software and the subroutines for GLMs are well developed
as, for example, the {\sf glm()} function, which is used to fit GLMs,
specified by giving a symbolic description of the linear predictor and a
description of the error distribution. Our Monte Carlo
simulation experiments indicated that the Wald test should not be
considered for testing hypotheses in GLMs when the sample is
small or of moderate size, however, the subroutines
of standard statistical softwares use the Wald test statistic to make
inference in this class of models. It reveals indeed the necessity of a package that contemplates
the corrected tests considered in this paper.

\section*{Acknowledgments}

We gratefully acknowledge grants from CNPq (Brazil), FAPESP (S\~ao Paulo, Brazil) and FACEPE (Pernambuco, Brazil).

\end{document}